\documentclass[twocolumn,
preprintnumbers,amsmath,amssymb] {revtex4}  

\usepackage{graphicx}
\usepackage{dcolumn}
\usepackage{bm}

\newcommand{\be}{\begin{equation}} \newcommand{\beq}{\begin{equation}}  
\newcommand{\ee}{\end{equation}}  \newcommand{\eeq}{\end{equation}}  
\newcommand{\bea}{\begin{eqnarray}}  
\newcommand{\eea}{\end{eqnarray}}  
\newcommand{\ba}{\begin{array}}  
\newcommand{\ea}{\end{array}}

\def\slash#1{#1\!\!\!/\!\,\,}  
\newskip\humongous \humongous=0pt plus 1000pt minus 1000pt

\newif\ifdtup

\def\oldreffmt#1{\rlap{[#1]} \hbox to 2\parindent{}}

\def\figfmt#1{\rlap{Figure {#1}} \hbox to 1in{}}  
  
\begin{document}

\preprint{FERMILAB-Pub-14-313-T}

\vskip 0.4in
\title{\Large Composite Vectorlike Fermions}

\author{ Bogdan A. Dobrescu and Christopher T. Hill}




\begin{abstract}

\vskip -0.3cm

\centerline{\it Theoretical Physics Department, Fermilab, P.O. Box 500, Batavia, Illinois 60510, USA}

\vskip 0.4cm

\centerline{September 4, 2014} 
\vskip 0.5cm

\normalsize
We study a dynamical mechanism that generates a composite vectorlike fermion, formed by the binding of an $N$-tuplet of elementary 
chiral fermions to an $N$-tuplet of scalars. Deriving the properties of the composite fermion in the large $N$ limit, 
we show that its mass is much smaller than the compositeness scale when the binding coupling is near a critical value. 
We compute the contact interactions involving four composite fermions, and find that their coefficients scale as $1/N$.
Physics beyond the Standard Model may include composite vectorlike fermions arising from this mechanism.
\end{abstract}


\maketitle

\section{
\large  
I\lowercase{ntroduction}}

All known elementary fermions are chiral, so that their masses 
arise as a consequence of electroweak symmetry breaking.
By contrast, vectorlike fermions have the
same gauge charges for left- and right-handed components, so that their Dirac mass terms
are present from the outset in the Lagrangian. Thus, it is natural 
to expect that vectorlike quarks or leptons, if they exist, are heavier than the Standard Model (SM) fermions. 
Vectorlike fermions are the subject of intensive searches at the LHC \cite{Chatrchyan:2013uxa}.

Vectorlike fermions are a key element within various strongly coupled theories 
for physics beyond the Standard Model (SM). Amongst these are 
the Top-seesaw theory \cite{Dobrescu:1997nm} where a 
composite Higgs boson \cite{Dobrescu:1999gv,Cheng:2013qwa} arises from the binding of a vectorlike quark to the top quark, 
models where the Higgs doublet is a pseudo-Nambu-Goldstone boson (pNGB) \cite{Agashe:2004rs}\cite{Cheng:2013qwa}, 
models of extra dimensions with bulk fermions \cite{Ponton:2012bi,Hill:2002ap}, and Little Higgs models \cite{ArkaniHamed:2001nc} where a vectorlike quark cancels the quadratic divergence due to the top quark.
In some of these theories the vectorlike fermions are thought to be bound states, but usually there is no precise description of what are their constituents,
or what is the binding interaction responsible for these  composite fermion fields.

Here we study a dynamical model of composite vectorlike fermions.  
We start with a dimension-6 interaction between a complex scalar and an elementary chiral fermion, both transforming in the fundamental representation of a global $SU(N)$ symmetry.
The dimension-6 interaction may be induced by heavy gauge boson exchange, 
similarly to coloron exchange  \cite{topcolor} in top condensation models \cite{Nambu, BHL}\cite{Hill:2002ap}.
Factorizing the interaction into spin-1 auxilliary fields, which at low energy acquire kinetic terms, we find that  
an $SU(N)$-singlet  composite Dirac fermion forms; its right-  (left-) handed component is an $s$-  ($p$-) wave bound state of the elementary fermion and scalar.
We solve this ``Composite Vectorlike Fermion" (CVF) model in the large $N$ limit, deploying the ``block-spin'' renormalization group (RG) \cite{BHL}. 
A model of this type was considered long ago 
\cite{Suzuki:1991kh} for describing the SM quarks and leptons as composite fields. 

Near a certain critical coupling the Dirac bound state becomes much lighter than the compositeness scale.
Although the Dirac mass vanishes at criticality, there is no associated chiral symmetry in this limit, 
due to the asymmetry between the dimension-4 and -5 operators producing the $s$- and $p$- wave
components, respectively.  

Large contact interactions have often been suspected of being associated with fermion compositeness \cite{EichtenLanePeskin}.
Using the large-$N$ limit of the CVF model, we compute the effective interactions involving four composite fermions, and find indeed large coefficients;
however, these coefficients scale as $1/N$, and thus become perturbative for very large $N$.

The CVF model has features in common with  the Nambu-Jona-Lasino (NJL)
model \cite{NJL}, which describes a spin-$0$ bound state,
composed of a right-handed fermion, and a left-handed anti-fermion.  The attractive interaction
which drives the formation of the bound state is
a chirally-invariant 4-fermion interaction, which may
be viewed as the relic of a coloron exchange \cite{topcolor}.
The leading effects of the interaction can be treated in large-$N$ approximation
and generate a composite scalar whose mass depends
upon a dimensionless coupling.  As this
approaches a critical value from below, the composite field becomes lighter, 
approaching masslessness.  Above critical coupling the bound state acquires
a vacuum condensate, the fermions acquire masses $m_f$ and pNGBs appear. 
There is also a Higgs boson of mass $2m_f$  in the broken phase in the large-$N$ approximation \cite{Nambu}.
The analysis can be improved by use of the block-spin RG \cite{BHL}.

Although in the minimal CVF model the composite fermion is a gauge singlet, it is easy to give its constituents charges under the SM gauge group and to obtain 
composite vectorlike quarks or leptons, as often employed in theories  beyond the SM.
We view this CVF model as a ``dynamics'' which can serve as the kernel of various composite models of fermions, including 
partially-composite SM quarks and leptons \cite{Kaplan:1991dc} and 
descriptions of heavy-heavy-light baryons.


\section{\large C\lowercase{omposite fermion model}}
\label{sec:model}

Consider a chiral fermion, $\psi_L$ and a complex scalar, $\phi$.
which transform in the fundamental representation of a global $SU(N)$ symmetry.   
We postulate a Lagrangian of the form
\beq
{\cal{L}} = {\cal{L}}_0 + {\cal{L}}_{\rm int}  ~~,
\eeq
where the free-fields Lagrangian is 
\beq
{\cal{L}}_0 = i\overline{\psi}_L \slash{\partial} \psi_L 
            + \partial_\mu \phi^\dagger \partial^\mu \phi 
	    - M_\phi^2 \phi^\dagger  \phi    ~~,
\eeq
and the interaction terms are
\beq
\label{eq3}
{\cal{L}}_{\rm int}=
  -\frac{g^2}{\Lambda^2}   \left( \, \overline{\psi}_L 
          \gamma^\mu T^a \psi_L  \right) \left(
           i \phi^\dagger  
  \stackrel{\leftrightarrow}{\partial}_\mu  T^a \phi \right) ~~.
\eeq
$T^a$ are the generators of $SU(N)$, and for the moment $\Lambda$ is a momentum space cut-off
on the loop integrals, with $M_\phi \ll \Lambda$. The notation in Eq.~(\ref{eq3}) is chosen to
suggest that this term may be  generated by the exchange of a gauge  boson of coupling $g$ and mass $\Lambda$.

Using the color Fierz identity to leading order in $1/N$,
\beq
T^a_{ij}  \, T^a_{k\ell }
= \frac{1}{2} \left( \delta_{i\ell }\delta_{k j} - 
\frac{1}{N} \delta_{ij }\delta_{k \ell } \right)
\approx
\frac{1}{2} \delta_{i\ell }\delta_{k j }  ~~,
\eeq
Eq.~(\ref{eq3}) becomes
\beq
\hspace*{-0.2cm}
{\cal{L}}_{\rm int} \approx \frac{ig^2}{2\Lambda^2}  \left[ \, \overline{\psi}_L \phi \right] \! \left[ ( \slash\partial \phi^\dagger)  \psi_L   \right]  + {\rm H.c.}
\eeq
where fields written within a pair of brackets, $[...]$, have their
$SU(N)$ indices contracted together.

The interaction term can be factorized \cite{Suzuki:1991kh} by introducing a static
$SU(N)$-singlet Dirac  fermion, $\chi$, as follows:
\beq
\label{eq6}
{\cal{L}}_{\rm int} = \widetilde{\cal{L}}_{\phi\psi\chi} -  \Lambda \overline{\chi}\chi  + O(1/N)  ~~,   
\eeq
where 
\beq
\widetilde{\cal{L}}_{\phi\psi\chi} =  i  \frac{g}{\Lambda}           \overline{\psi}_L          (\slash\partial \phi)\chi_L
-  \frac{g}{2}\, \overline{\chi}_R{\phi}^\dagger \psi_L  + {\rm H.c. }       
\label{eq:trilinear}
\eeq
Integrating out $\chi$ we recover Eq.~(\ref{eq3}) in the large-$N$ limit. 
Therefore, we can view Eqs.~(\ref{eq6}) and (\ref{eq:trilinear}) as the 
form of the interaction at the scale $ \Lambda$, which can then be evolved  downward in scale to $\mu \ll\Lambda$.

Note that the choice of the couplings and mass of $\chi$ at this stage
is somewhat arbitrary, due to our freedom to rescale $\chi_L$ or
$\chi_R$.  Momentarily, the $\chi$  fields will develop
kinetic terms which will ultimately be canonically normalized, fixing the coupling normalizations.

\vspace{.2cm}

\section{\large L\lowercase{ow-energy effective theory}}
\label{sec:effective}

The low-energy effective Lagrangian may be derived most expeditiously by 
means of the block-spin RG.
In the case of the NJL model, the block-spin RG  has been developed in \cite{BHL}. 
Here we view Eq.~(\ref{eq6}) as the effective Lagrangian of the theory at a
mass scale $\Lambda$.  To derive the effective
Lagrangian at a lower scale $\mu$, where
$\Lambda > \mu > M_\phi$,  we integrate out the field modes of momenta $\Lambda \geq k \geq \mu$. 
For the factorized CVF model of Eq.~(\ref{eq6}) this yields the effective Lagrangian at scale $\mu$:
\beq
{\cal{L}} (\mu) =  \! {\cal{L}}_{0} + \widetilde{\cal{L}}_{\phi\psi\chi \!} 
+   Z_L\overline{\chi}_L{i\slash{\partial}} \! \chi_L
        +Z_R\overline{\chi}_R {i\slash{\partial}} \! \chi_R - \widetilde{m}_\chi \overline{\chi} \chi  ~ , 
        \label{eq:Lmu}   
\eeq
where we neglected operators of dimension 6 or higher.
The $\chi_R$ and $\chi_L$ fields have acquired kinetic terms from the first diagram of
Fig.~1 (the leading-$N$ contribution), and thus have become dynamical fields.
Their wave-function renormalizations are given for $M_\phi \ll \mu$  by
\bea
\label{Z_R}
&& Z_R = \frac{\kappa N}{32\pi}\ln\left(\frac{\Lambda^2}{\mu^2} \right)  ~~,
\nonumber \\ [2mm]
&& Z_L = \frac{\kappa N \eta}{4\pi}\left( 1- \frac{ \mu^2}{\Lambda^2} \right)  ~~.
\label{eq:ZL}
\eea
The matching coefficient $\eta$, of order one, depends on the procedure of cutting off the quadratic  divergence in $Z_L$; 
integrating over the loop momentum $k^\mu$ (indicated in Fig.~1), with integration limits  $\mu \leq |k| \leq \Lambda $, gives $\eta=1$.
We defined the coupling constant
\beq
\kappa \equiv \frac{g^2}{4 \pi} ~~ . 
\eeq
The Dirac mass of Eq.~(\ref{eq:Lmu}), arising from the second diagram of Fig.~1, is given in units of $\Lambda$ by
\beq
\frac{ \widetilde{m}_\chi }{\Lambda} =  1- \frac{\kappa N}{8 \pi}\left[ 1- \frac{\mu^2}{\Lambda^2}  +O\!\left( \frac{M_\phi^2}{\Lambda^2} \ln \!\Big(\frac{\mu^2}{\Lambda^2} \Big)\! \right) \,\right] ~~.
\eeq

\begin{figure}[t]
\begin{center}
\hspace*{-0.25cm}
\includegraphics[trim = 35mm 177mm 40mm 28mm, clip, height=4.6cm]{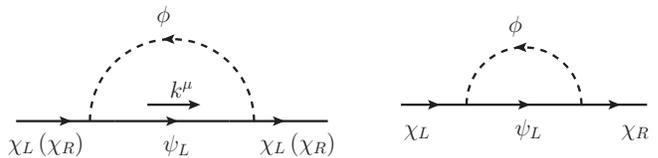}
\vspace*{-2.7cm}
\caption{Large-$N$ contribution to the  wave-function renormalizations $Z_L$ and $Z_R$ (left diagram), and to the Dirac 
$\chi$ mass (right diagram). Integration over the loop momentum $k^\mu$, 
with $|k|$ cut-off at $\Lambda$, gives $\eta=1$ in Eq.~(\ref{eq:ZL}).
}
\end{center}
\label{feyn_diagrams}
\end{figure}

It is useful to compare the NJL model side-by-side with the present scheme. 
First, $Z_R$ is a conventional log-running result for an NJL type theory,
{\it e.g.} of a composite Higgs \cite{BHL}. 
In NJL an auxiliary  scalar field, $H$ (a Higgs field), 
is introduced to factorize a 4-fermion interaction, 
in analogy to the auxiliary fermion  field $\chi$ in our CVF model.
Applying the block-spin RG to the NJL model, 
one obtains an induced kinetic term
for $H$ with a logarithmic wave-function renormalization
constant $Z_H$, in analogy to $Z_R$ in Eq.~(\ref{Z_R}).

The induced kinetic terms in NJL always vanish as $\mu \rightarrow \Lambda$
and this is normally referred to as the ``compositeness matching
conditions.''  The $\chi_R$ kinetic term is generated by a
dimension-4 interaction, $\overline{\chi}_R{\phi}^\dagger \psi_L$,
so that the dynamical $\chi_R$ field may be viewed as an
$s$-wave bound state of $\phi$ with $\psi$, just as
$H$ is an $s$-wave bound state of $\bar{t} t$ in the top-condensation theory \cite{BHL}.

Second, the wave-function renormalization constant for the composite $\chi_L$ field,
$Z_L$, has the behavior in the block-spin RG of quadratic running in scale $\mu$,
analogous to the Higgs boson mass in top condensation.
 It arises from two insertions of the dimension-5 vertex 
$\overline{\psi}_L (\slash \partial \phi)\chi_L$ (see the first diagram of Fig.~1), 
hence the quadratic divergence occurs up to the cut-off
$\Lambda$.  The exact result for $Z_L$ is sensitive to how the theory is cut-off in detail.
A precise determination of the coefficient $\eta$ is really a matching condition
of the low-energy theory onto a realistic high energy theory, such as a
coloron model. This is typical for a quadratic or higher divergence, since the
divergence is emphasizing the short-distance limit of the
theory.  This means that $\chi_L$ enters the theory as a $p$-wave
bound state, and the quadratic vanishing of $Z_L$ as $\mu \rightarrow
\Lambda$ is indicative of the faster short-distance vanishing of the wave-function for the 
$p$-wave.

Finally, the Dirac mass $ \widetilde{m}_\chi$ is also a quadratic divergence
associated with the second diagram of Fig.~1.  The analogous quantity in
conventional NJL is the squared mass, $M^2_H$ for
the composite $H$ field.  As $\mu \rightarrow \Lambda$
we see that $ \widetilde{m}_\chi \rightarrow \Lambda$, {\it i.e.},
we map back onto the original  factorized effective Lagrangian.

\vspace{.2cm}

\section{\large P\lowercase{hysical fields and critical behavior}}
\label{sec:critical}

We can pass to canonical normalizations for the $\chi$ fields
by the scaling redefinitions
\beq
\chi_R \rightarrow \sqrt{Z_R} \, \chi_R \;\; ,\;\;  \chi_L \rightarrow \sigma  \sqrt{Z_L}\, \chi_L ~,
 \label{eq:redef}
\eeq
where $\sigma = \pm 1$ is introduced for later convenience. The effective Lagrangian at the scale $\mu$ becomes
\beq
{\cal{L}}(\mu)=  {\cal{L}}_{0} + {\cal{L}}_{\phi\psi\chi} +
         \overline{\chi} {i\slash{\partial}}\chi
                 - m_\chi \overline{\chi} \chi 
\label{eq:effective}
\eeq
with interaction terms
\beq
{\cal{L}}_{\phi\psi\chi} =  \!
 \frac{ i \sigma  g}{\Lambda \sqrt{Z_L}}  
         \overline{\psi}_L 
         (\slash\partial \phi)\chi_L
- y_\chi \, \overline{\chi}_R{\phi}^\dagger \psi_L 
+ {\rm H.c. }       
\label{eq:trilinear-rescaled}
\eeq
The Yukawa coupling from the second term of  ${\cal{L}}_{\phi\psi\chi}$ is given by 
\beq
y_\chi = \frac{g}{2 \sqrt{Z_R}} = \frac{4\sqrt{2}\, \pi }{\sqrt{N \ln \left( \Lambda^2/\mu^2 \right)} } ~~.
\eeq 
This is the coupling of the $\chi_R$ fermion to its constituents, 
so it is large, blowing up as expected at the compositeness scale $\Lambda$. 
The Yukawa coupling becomes perturbative only at scales $\mu \ll 10^{-5} \Lambda$, since the expansion parameter is
$y^2 N/(4\pi) = 4\pi /\ln (\Lambda/\mu)$.

The physical $\chi$ mass, which enters Eq.~(\ref{eq:effective}), evaluated at the scale $\mu \gg M_\phi$ is
\bea
\hspace*{-1.9cm}
&& \hspace*{-.9cm} m_\chi =  \sigma \frac{\widetilde{m}_\chi}{\sqrt{Z_L Z_R}} 
 \\ [1mm]
&& \hspace*{-.3cm} =   \sigma\Lambda  
\left( \frac{\kappa_c }{\kappa}  - 1 +  \frac{\mu^2}{\Lambda^2}\right)
 \left[\frac{\eta}{2}\! \left( \!1\! - \!\frac{\mu^2}{\Lambda^2} \!\right)\ln\!\left( \frac{\Lambda^2}{\mu^2 } \right)\!\right]^{-1/2} 
 \!\!\! ,
\nonumber
\eea
where we kept the dependence on the matching coefficient $\eta$  introduced in Eq.~(\ref{eq:ZL}).
The behavior of this Dirac mass as $\mu \rightarrow 0$ is controlled by
the critical coupling constant
\beq
\kappa_{c}  \equiv \frac{g^2_c}{4 \pi} = \frac{8\pi}{N}  ~~.
\eeq
For the weak-coupling case, $\kappa \ll  \kappa_c$, the composite fermion decouples ($ m_\chi >  \Lambda   $),  
so the low-energy physics includes just $\psi_L$ and  $\phi$.  

For $\kappa < \kappa_c $ with $\kappa_c - \kappa \ll \kappa$, the Dirac mass 
becomes much smaller than $\Lambda$ for $\mu \ll \Lambda$. This is not surprising, since the analogue 
behavior occurs in the conventional NJL model.  We are essentially tuning
a large hierarchy as $\kappa \rightarrow \kappa_c$, and an approximately
scale-invariant theory near the critical coupling.

In a conventional NJL  model the behavior of the bound state boson
mass, $M^2_H$, as 
$\mu \rightarrow 0$ determines the criticality of the model. 
If $M^2_H <0$ for $\mu \rightarrow 0$, then the composite Higgs field acquires a VEV. 
The critical coupling of NJL is that value for which $M_H^2 =0$ as $\mu \rightarrow 0$.

In the CVF model at the exact critical value we have $m_\chi\rightarrow 0$.
This is not, however, a chirally invariant theory, since 
the original interactions of the full theory violate the chiral
symmetry of $\chi$.  Nonetheless, the effective Lagrangian
does contain the massless fields $\chi_R$ and $\chi_L$.
If one naively discards the ``irrelevant'' operator of $\cal L_{\phi\psi\chi}$, {\it i.e.}, the first term 
of Eq.~(\ref{eq:trilinear-rescaled}), in this limit, then we would have a true chiral fermion.

For $\kappa > \kappa_c$ the Dirac mass becomes negative at scales $\mu \ll \Lambda$. 
At the exact scale $\mu_0$ where $m_\chi(\mu_0) = 0$, 
\be
\mu_0 = \Lambda \left( 1 - \frac{\kappa_c}{\kappa} \right)^{1/2} ~~,
\ee
the full chiral symmetry of $\chi$, $U(1)_L \times U(1)_R$, is broken 
only by the $\cal L_{\phi\psi\chi}$ interactions of Eq.~(\ref{eq:trilinear-rescaled}). It is thus possible to
redefine the $\chi_L$ and $\chi_R$ fields by arbitrary complex phases, with the only effect being a change in the relative phase
between the two terms of $\cal L_{\phi\psi\chi}$.
In order to keep $m_\chi (\mu) \ge 0 $ at all scales, we redefine  $\chi_L \to -\chi_L$ at scales $\mu \leq \mu_0$.
In other words, the sign introduced in Eq.~(\ref{eq:redef}) is $\sigma =+1$ for $\mu > \mu_0$, and $\sigma =-1$ for $\mu \leq \mu_0$.
This means that the RG evolution of  $\sigma (\mu)$  
is proportional to the step function centered at $\mu_0$. 
The dependence of $m_\chi $ on $g/g_c$ is shown in Fig.~2.  

\begin{figure}[t!]
\begin{center}
\includegraphics[width=0.45 \textwidth ]{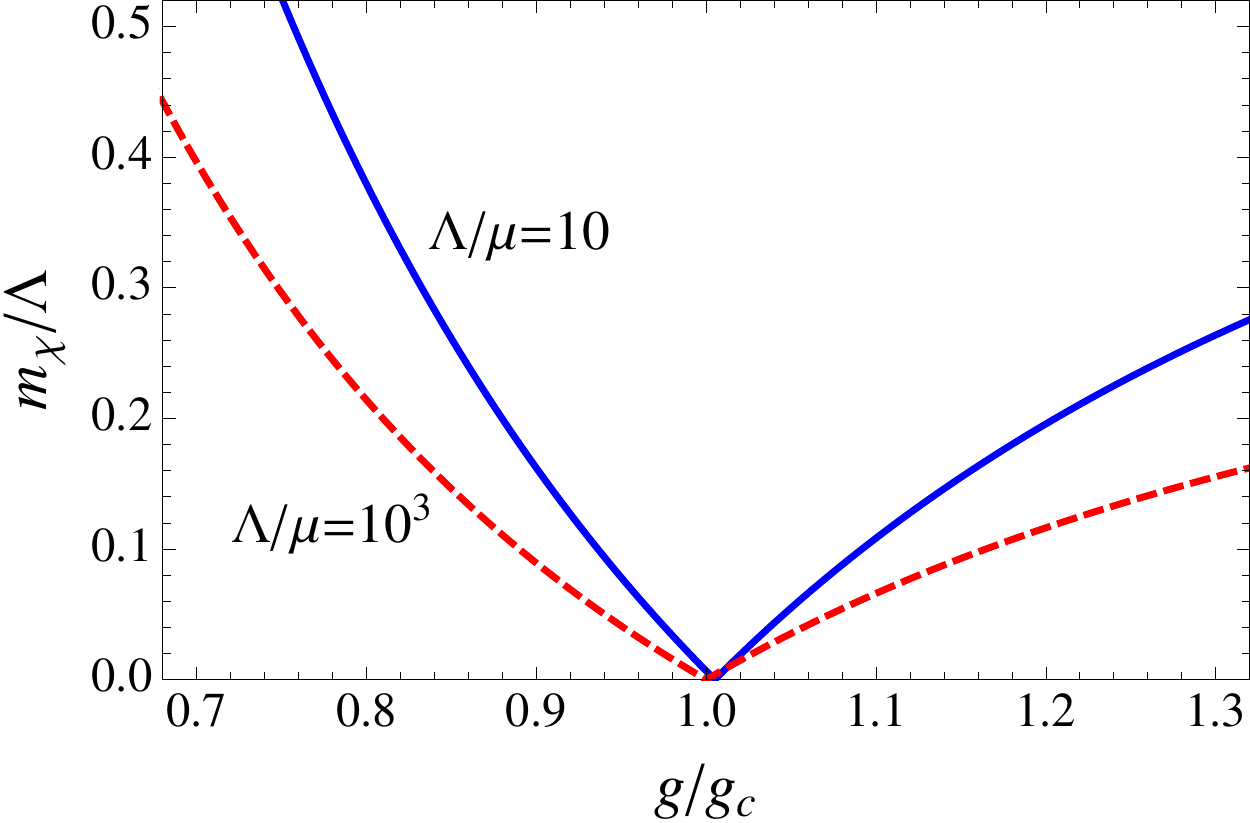}
\caption{ Mass of composite fermion $\chi$ (relative to the compositeness scale $\Lambda$), at the scale $\mu$, as a function of the binding coupling $g$ (in units of the critical coupling $g_c = 4\pi \sqrt{2/N}$) for $\Lambda/\mu = 10$ or $10^3$, and matching coefficient $\eta = 1$.}
\label{fig:mchi}
\end{center}
\end{figure}

\vspace{.2cm}

\section{\large F\lowercase{our-fermion interactions}}
\label{sec:four-fermion}

The Lagrangian at scale $\mu$, ${\cal{L}} (\mu)$ includes operators of dimension-6 or higher, not shown in Eq.~(\ref{eq:Lmu}).
Particularly important among those are  4-$\chi$ terms. Large contact interactions of this type have often been suspected of being associated with fermion compositeness \cite{EichtenLanePeskin}.
In our dynamical model, the coefficients of the operators involving four composite fermions  can be computed in the large $N$ limit.

A peculiar 4-$\chi$ term is the one involving only right-handed fields,
because it is generated only by the dimension-4 vertex involving the massless fermion $\psi_L$ (see the second term of ${\cal{L}}_{\phi\psi\chi} $). 
As a result, its coefficient is infrared divergent, instead of being suppressed by $\Lambda^2$. 
Performing the loop integral shown in Fig.~3, we find the following operator involving $\chi_R$ (with canonically normalized kinetic term):
\beq
\frac{-8 \pi^2}{N (\mu^2 + M_\phi^2) } 
\Big[  \ln \Big(\frac{\Lambda^2}{\mu^2+ M_\phi^2}\Big)\! \Big]^{-2}  
\, (\overline{\chi}_R\gamma_\mu\chi_R)^2 ~,
\eeq
for $\Lambda^2 \gg \mu^2, M_\phi^2$.
Note that the infrared divergence is cut-off either by $M_\phi$ or by the scale $\mu$ where the coefficient is evaluated (a physical value for that is $\mu = m_\chi$).

The other 4-$\chi$ operators are given by 
\be
\frac{8\pi^2}{N \eta  \Lambda^2} \left[ 2 (\overline{\chi}\chi)^2 
+  (\overline{\chi}_L\gamma^\mu\chi_L)(\overline{\chi}_R\gamma_\mu\chi_R)
- \frac{\eta^\prime}{4\eta }  
(\overline{\chi}_L\gamma_\mu\chi_L)^2  \right]
 \ee
Here we have included a matching coefficient $\eta'$ (of order one) in the quadratically
divergent loops, in addition to the factor of $\eta$ from $Z_L$. Cutting off the loop integral at $\Lambda$ gives $\eta' = 1$.

The chiral symmetry breaking in these terms is a consequence of the 
structure of  ${\cal{L}}_{\phi\psi\chi} $, 
shown in Eq.~(\ref{eq:trilinear-rescaled}).
As expected, the coefficients of these contact terms in units of $\Lambda^{-2}$ include a large factor of order $(4\pi)^2$. Nevertheless, for very large $N$ these coefficients become perturbative.  

\begin{figure}[t]
\begin{center}\hspace*{-0.5cm}
\includegraphics[trim = 25mm 177mm 40mm 28mm, clip, height=5.3cm]{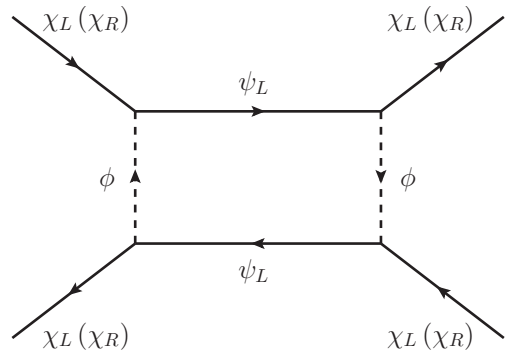} 
\caption{Large-$N$ contribution to the  4-$\chi$ operators. The vertices involving $\chi_R$ ($\chi_L$) are due to a Yukawa coupling (dimension-5 operator), as given in Eq.~(\ref{eq:trilinear-rescaled}).}
\end{center}
\label{diagrams4chi}
\end{figure}

\vspace*{.5cm}

\section{\large V\lowercase{ectorlike quarks and leptons}}
\label{sec:realistic}

So far we have shown how a Dirac fermion is produced by the binding of a scalar $N$-tuplet $\phi$ to a chiral fermion $N$-tuplet $\psi_L$. Let us now 
see how to use this mechanism to generate vectorlike fermions that carry some $SU(3)_c\times SU(2)_W \times U(1)_Y$ gauge charges, as required in various extensions of the SM.
If there is a single $\psi_L$, then the gauge anomaly cancellations are not consistent with it carrying hypercharge or transforming in complex representations of 
$SU(3)_c\times SU(2)_W $. The scalar $\phi$, however, can carry any gauge charges.

A frequently encountered vectorlike fermion \cite{Dobrescu:1997nm,Agashe:2004rs,ArkaniHamed:2001nc} 
transforms as $(3,1,-1/3)$ under the SM gauge group. This can be a bound state of a $\psi_L$ that is a singlet under the SM group,
and a complex scalar $\phi$ that transforms as $(3,1,-1/3)$. Note that $\psi_L^c$ can be referred to as a ``right-handed neutrino", while $\phi$
can be identified with a ``right-handed squark" of the supersymmetric SM.

If $\psi_L$ is part of a larger set of chiral fermions, each participating together with a scalar $\phi$ in dimension-6 operators of the type (\ref{eq3}), then 
there will be a composite vectorlike fermion for each of the elementary chiral fermions. One could imagine applying this set of ideas to the SM model
in order to explain the quark and lepton mass hierarchies \`{a} la Froggatt-Nielsen \cite{Froggatt:1978nt}.

As we alluded earlier, the UV completion responsible for the dimension-6 operator (\ref{eq3}) may simply be a heavy gauge boson. If the 
 $SU(N)$ symmetry is gauged and spontaneously broken, the one-gauge-boson exchange between $\psi_L$ and $\phi$ induces  operator (\ref{eq3}).
Note that the $SU(N)$ symmetry would be anomalous unless additional chiral fermions transform under $SU(N)$. 
An alternative is that the global $SU(N)$ group is an accidental symmetry arising from a spontaneously-broken  gauge group under which $\psi_L$ 
transforms in a real representation. For example, $\psi_L$ and $\phi$  may belong to the adjoint representation of an $SU(N_0)$ gauge group, with $N_0^2- 1 = N$.
Whether the gauge symmetry that provides the binding is $SU(N)$  or $SU(N_0)$, there are bound states in addition to the vectorlike fermion, {\it e.g.},
 a $\phi\phi$ composite scalar, that remain to be studied.

\section{\large C\lowercase{onclusions}}
\label{sec:conc}\setcounter{equation}{0}

We have shown that a composite Dirac fermion arises as a bound state of a complex scalar and a chiral fermion, both belonging to the $N$ representation of a global $SU(N)$ group.
The binding is provided by an attractive dimension-6  interaction. The origin of this non-confining interaction could be an asymptotically-free  gauge group that is 
spontaneousy broken near the scale where it becomes strongly coupled.

An advantage of compositeness formulated in this way is that, at least in the large-$N$ approximation, one has control over the 
model and such things as the Dirac mass of the composite fermion and 4-fermion contact terms may be computed. 
The Dirac mass vanishes when the coefficient of the
dimension-6  interaction is given by $-g_c^2/\Lambda^2$, where $g_c = 4\pi \sqrt{2/N}$ is the critical
coupling and $\Lambda$ is the compositeness scale. We have found that although the induced 4-fermion interactions have typically large coefficients, 
they are suppressed by $1/N$ and become perturbative for $N \gtrsim O(8\pi^2)$.

We have assumed that the scalar $N$-tuplet $\phi$ has a mass much smaller than the fermion compositeness scale. This hierarchy could be enforced by
${\cal N} = 1$ supersymmetry, since the elementary chiral fermion and $\phi$ may form a chiral superfield. In that case, if the dimension-6 operator is induced from a heavy 
gauge boson exchange, then additional higher-dimension operators arising from gaugino exchange need to be analyzed.
Alternatively, the $M_\phi \ll \Lambda$ hierarchy may be natural if $\phi$ is a pNGB, perhaps within some underlying strongly-coupled theory.

Vectorlike fermions are deployed in a number of models in the literature.  Many of these models could be adapted
along the present lines to engineer vectorlike fermions as composites.
For example, it is not hard to imagine composite Higgs models \cite{Dobrescu:1999gv,Cheng:2013qwa} based on the Top-seesaw mechanism  \cite{Dobrescu:1997nm} in
which the $\chi_{L}$ and $\chi_R$ top partners arise as composite states as described here.

\vspace*{.7 cm}

{\bf Acknowledgments}: 
We would like to thank Bill Bardeen for insightful conversations. Fermilab is operated by Fermi Research Alliance, LLC under Contract No. 
DE-AC02-07CH11359 with the United States Department of Energy.

\vspace{ 1 cm}


\end{document}